\documentclass[9pt]{article}
\usepackage{spconf,amsmath,amsfonts,graphicx}

\usepackage{hyperref}
\usepackage{bm}

\title{Attentive Filtering Networks for Audio Replay Attack Detection}
\name{Cheng-I Lai$^{1,2}$, Alberto Abad$^{2,3}$, Korin Richmond$^2$, Junichi Yamagishi$^{2,4}$, Najim Dehak$^1$, Simon King$^2$}
\address{$^1$Center for Language and Speech Processing, Johns Hopkins University, USA\\
$^2$Centre for Speech Technology Research, University of Edinburgh, UK\\
$^3$INESC-ID / Instituto Superior T\'ecnico, University of Lisbon, Portugal\\
$^4$Digital Content and Media Sciences Research Division, National Institute of Informatics, Japan}
\begin{document}
\ninept
\maketitle
\begin{abstract}
An attacker may use a variety of techniques to fool an automatic speaker verification system into accepting them as a genuine user.  Anti-spoofing methods meanwhile aim to make the system robust against such attacks. The ASVspoof 2017 Challenge focused specifically on replay attacks, with the intention of measuring the limits of replay attack detection as well as developing countermeasures against them. In this work, we propose our replay attacks detection system - Attentive Filtering Network, which is composed of an attention-based filtering mechanism that enhances feature representations in both the frequency and time domains, and a ResNet-based classifier. We show that the network enables us to visualize the automatically acquired feature representations that are helpful for spoofing detection. Attentive Filtering Network attains an evaluation EER of 8.99$\%$ on the ASVspoof 2017 Version 2.0 dataset. With system fusion, our best system further obtains a 30$\%$ relative improvement over the ASVspoof 2017 enhanced baseline system. 

\end{abstract}
\section{Introduction}
\label{sec:intro}
Automatic speaker verification (ASV) systems have become increasingly widespread in recent years with the advent of voice assistant and smart home devices. However, these systems may be vulnerable to presentation attacks, which are also known as spoofing attacks~\cite{wu2015spoofing}. One way an impostor might attempt to fool an ASV system into accepting them as a target speaker is by mimicking the voice characteristics of a genuine user. In order to ensure the continued reliability of ASV technology, it is necessary to develop countermeasures to protect against such spoofing attacks. There are four types of spoofing attack~\cite{wu2015spoofing}: Impersonation, Replay, Speech Synthesis, and Voice Conversion. The ASVspoof2017 Challenge is based on Replay attacks. The objective of the challenge is to develop countermeasures to defend against replay attacks, and to measure the limits of replay attack detection~\cite{delgado2018asvspoof}.

Previous work on the ASVspoof2017 Challenge data can generally be divided into three categories:  Gaussian Mixture Model (GMM) and i-vector based systems~\cite{delgado2018asvspoof,adiban2017sut}; deep neural network-based systems~\cite{lavrentyeva2017audio,valenti2018end,chen2017resnet}; and systems fusing the preceding models. In~\cite{delgado2018asvspoof}, Constant Q Cepstral Coefficient (CQCC) features together with a GMM classifier and i-vectors formed the enhanced baseline system for the Challenge. In~\cite{adiban2017sut}, thorough experiments were conducted on GMMs, i-vectors and Multi-Layer Perceptrons. The authors of~\cite{lavrentyeva2017audio} employed Light Convolutional Neural Networks (LCNN) and stacked LCNNs with a recurrent neural network (RNN). In~\cite{valenti2018end}, the authors developed an evolution RNN for spoofing detection. We also noticed recently published work~\cite{alam2018boosting,saranyadecision,sailor2018auditory,shim2018replay,suthokumar2018modulation} on Version 1.0 and 2.0 dataset, and we have included and compared their results in our experiments section.

The goal of the work here is to develop a deep learning system that utilizes discriminative features in both the time and frequency domain for spoofing detection. The motivation is that clues for spoofing attacks may be time varying and only partially observable, perhaps in the noise between spoken words or the high frequency components of speech for example. However, we are not sure where these clues might be embedded within the feature spaces. Therefore, we desire a system that automatically acquires and enhances discriminative time and frequency features that are helpful for the detection of spoofing attacks. We achieve this by designing an attention mechanism-based filter to cancel or enhance features prior to a ResNet-based classifier.

We took inspiration from three prior studies when designing the attention mechanism-based filter. Stimulated training~\cite{ragni2017stimulated} encourages activations to group in an interpretable way by superimposing a phone set during neural network acoustic model training. This inspired us to look at ways of applying an attention mechanism in model training. The convolutional attention network~\cite{lee2018convolutional} computes an attention matrix from speech spectrograms and embedded word sequences and multiplies it with the spectrogram prior to the classifier. This work inspired us to apply an attention mechanism prior to a classifier. The residual attention network~\cite{wang2017residual} applies a bottom-up feedforward process and top-down attention feedback.  This work inspired us to adopt similar processes within our filter. 

As for the classifier, while a ResNet has already been applied to the Challenge data~\cite{chen2017resnet}, we believed the network architecture may be further improved. Our ResNet-based classifier, which we term a Dilated Residual Network (DRN), uses convolution layers instead of fully connected layers, and we modify the residual units by adding a dilation factor. The filter and the classifier together compose our proposed method: Attentive Filtering Network (AFN).

The remainder of this paper is organized as follows: We first describe the Attentive Filtering Network in detail, including feature engineering, dilated residual network, attentive filtering, and optimization. Then, we briefly describe the three baseline systems that we re-implemented. This is followed by experimental setup, results and discussion. We end the paper with some concluding remarks.


\section{Attentive Filtering Network}
\label{sec:network}
\subsection{Feature Engineering}
\label{ssec:feature}
\begin{figure}[t]
  \centering
  \includegraphics[width=0.9\linewidth]{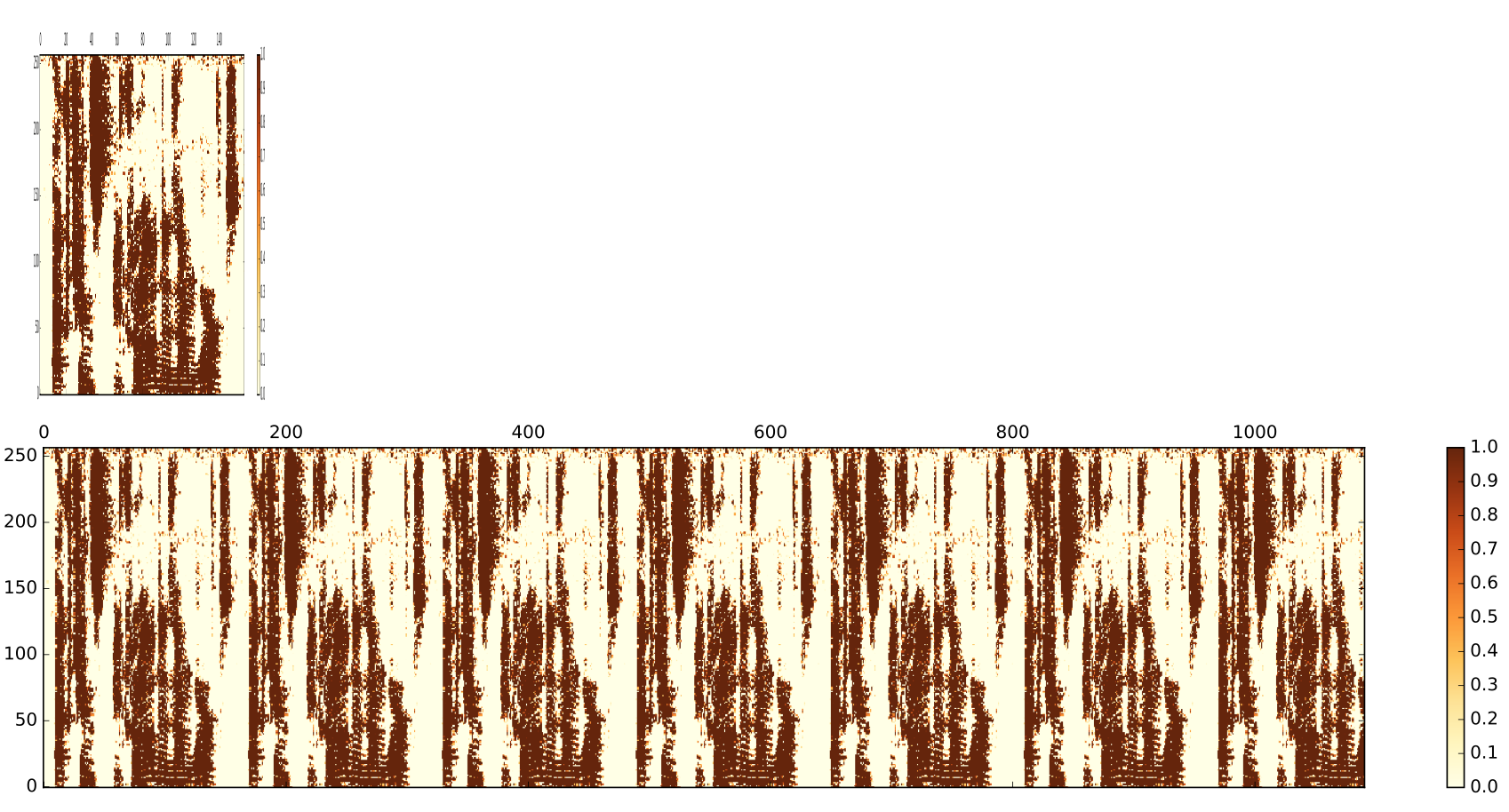}
  \vspace{-4mm}
  \caption{An illustration of unified time-frequency map creation by extending the log power magnitude spectra of all utterances to the length of the longest utterance. \textbf{Top} is the original spectra and \textbf{Bottom} is the unified time-frequency map.}
  \vspace{-4mm}
  \label{fig:feature-eng}
\end{figure}

We created a unified time-frequency map from log power magnitude spectra (logspec) obtained via Fast Fourier Transform as the input for our network. The dimension of logspec is 257. We kept all frames without applying voice activity detection, and applied mean normalization using a 3-second sliding window. The unified time-frequency map was created by extending all utterances to the length of the longest utterance by repeating their feature maps, illustrated in Figure~\ref{fig:feature-eng}. The resulting dimensionality of time-frequency maps for all utterances is 257 (frequency-domain) by 1091 (time-domain).  The benefits of this feature engineering approach are that there is no need for truncating features~\cite{lavrentyeva2017audio}, and since it is an utterance-level feature representation, there is no need for frame-level score combination. 

\subsection{Dilated Residual Network}
\label{ssec:DRN}
\begin{figure}[t]
  \centering
  \includegraphics[width=0.8\linewidth]{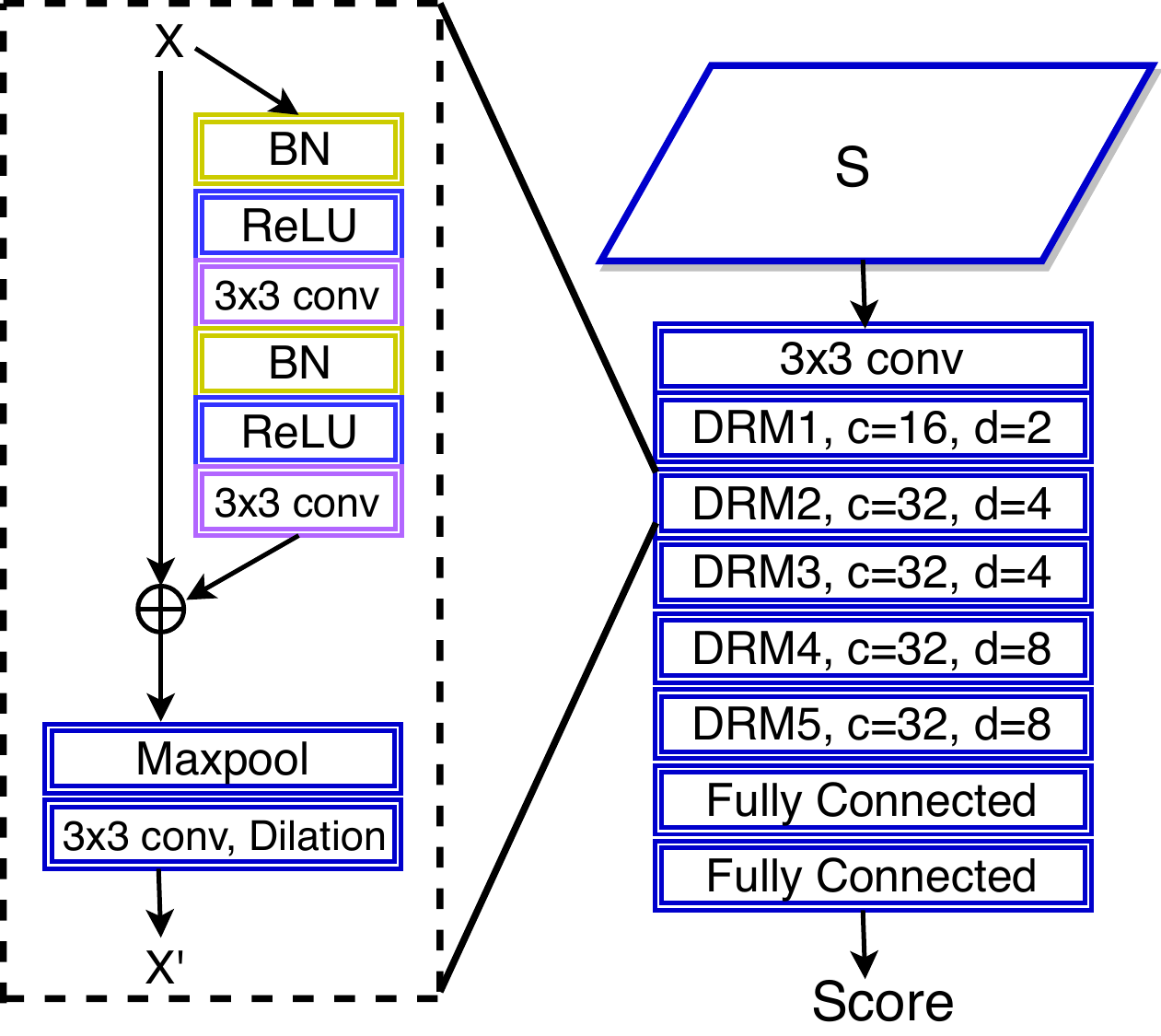}
  \vspace{-3mm}
  \caption{Illustration of (\textbf{Left}) Dilated Residual Module (DRM) and (\textbf{Right}) Dilated Residual Network (DRN). $S$ is the logspec input to the DRN, which is itself composed of five DRM blocks. Within each block, $c$ is the channel dimension (16 \& 32) and $d$ is the dilation rate (2, 4 \& 8). In addition to ReLU, we also experimented with ELU as an alternative activation function for the DRM.}
  \vspace{-1mm}
  \label{fig:drn-drm}
\end{figure}

A dilated residual network is composed of five Dilated Residual Modules (DRM), as shown in the right side of Figure \ref{fig:drn-drm}. Each DRM has a 3$\times$3 CNN based residual unit similar to \cite{he2016identity}, followed by a max-pooling layer and a dilated convolution layer, as illustrated in the left side of Figure \ref{fig:drn-drm}. 



The motivation for including a dilated convolution layer in the DRM arises from the observation that most of the previous ASVspoof 2017 Challenge-related research reported problems with generalizing to unseen conditions. In particular, the training set is small and the evaluation set contains very different conditions from those in the training and development sets. Therefore, prevention of overfitting is an important factor for obtaining good performance, and one idea to reduce the effects of overfitting in small datasets  without compromising model capacity is to include dilation in convolution. Dilated convolution operation $*_{d}$ is defined as~\cite{yu2015multi}:
\begin{equation}
  (F *_{d} G)(\mathbf{n}) = \sum_{\mathbf{m_{1}}+d\mathbf{m_{2}}=\mathbf{n}} F(\mathbf{m_{1}})G(\mathbf{m_{2}}), \forall \mathbf{m_{1}}, \mathbf{m_{2}},
\end{equation}
where $F$ is the feature map, $G$ is the kernel, $d$ is the dilation rate, $\mathbf{m_{1}}, \mathbf{m_{2}}$ and $\mathbf{n}$ are vectors. With dilated convolution, the network's receptive field grows exponentially with layer depth such that it integrates knowledge for the wider and global context~\cite{yu2015multi}. Max-pooling layers permit reduction of the spatial dimension of the feature maps. 


\begin{table}[tb]
\footnotesize
\caption{Configurations of the five Dilated Residual Modules in the Dilated Reisudal Network}
\vspace{2mm}
\label{fig:drn_summary}
\begin{tabular}{|c|c|c|c|c|c|}
\hline
Block & DRM1 & DRM2 & DRM3 & DRM4 & DRM5 \\ \hline
Convolution & 3$\times$3 & 3$\times$3 & 3$\times$3 & 3$\times$3 & 3$\times$3 \\ \hline
Dilation Size & 2 & 4 & 4 & 8 & 8 \\ \hline
Receptive Field & 7$\times$7 & 15$\times$15 & 23$\times$23 & 39$\times$39 & 55$\times$55 \\ \hline
Input Channels & 16 & 32 & 32 & 32 & 32 \\ \hline
Output Channels & 32 & 32 & 32 & 32 & 32 \\ \hline
\end{tabular}
\vspace{-2mm}
\end{table}

\subsection{Attentive Filtering}
\label{ssec:Attentive_filtering}
\begin{figure}[t]
  \centering
  \includegraphics[width=0.6\linewidth]{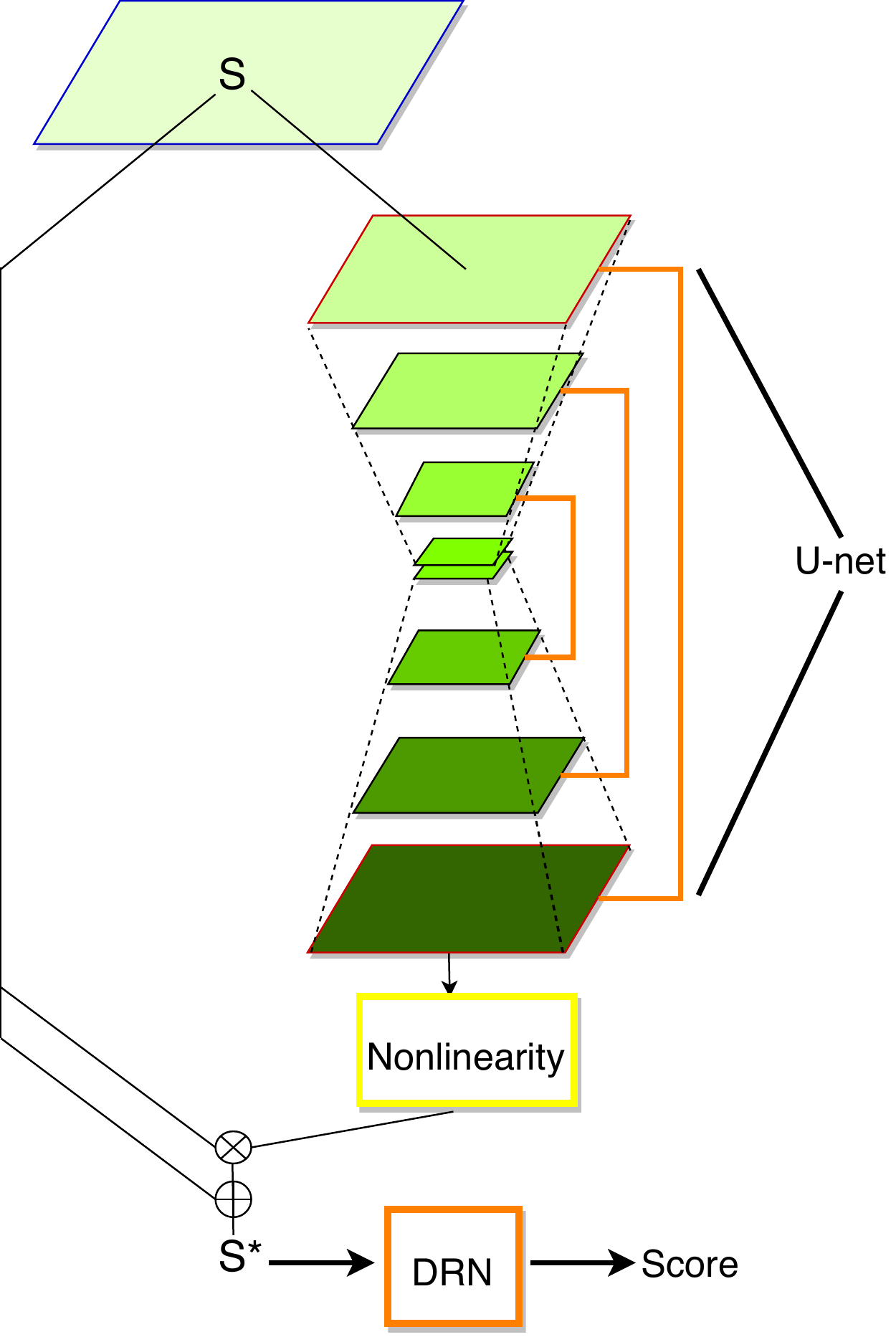}
  \vspace{-3mm}
  \caption{Illustration of Attentive Filtering. Input $\mathbf{S}$ is a feature map, and output $\mathbf{S}\mbox{*}$ is the new input for the DRN. We experimented with four nonlinear transforms: \textit{Sigmoid}, \textit{Tanh}, \textit{SoftmaxT} and \textit{SoftmaxF}. Here, \textit{SoftmaxT} means a \textit{softmax} operation in the time domain, and a \textit{SoftmaxF} operation means \textit{softmax} in the frequency domain. }
  \label{fig:af}
  \vspace{-5mm}
\end{figure}

Attentive Filtering (AF) accumulates discriminative features in frequency and time domains selectively. AF augments every input feature map $\mathbf{S}$ with an attention heatmap $\mathbf{A_{s}}$. The augmented feature map $\mathbf{S}\mbox{*}$ is then treated as the new input for the DRN as shown in Figure \ref{fig:af}. For $\mathbf{S},\mathbf{S}\mbox{*}\in\mathbb{R}^{F\times T}$, AF is described as: 
\begin{align}
  \mathbf{S}\mbox{*} = \mathbf{A_{s}} \circ \mathbf{S} \bm{+} \overline{\mathbf{S}}, \label{eq:attent}
\end{align}
where $F$ and $T$ are the frequency and time dimensions, $\circ$ is element-wise multiplication operator, $\bm{+}$ is element-wise addition operator, and $\overline{\mathbf{S}}$ is the residual $\mathbf{S}$. In this work, we set $\overline{\mathbf{S}} = \mathbf{S}$. To learn the attention heatmap, $\mathbf{A_{s}}$ contains similar bottom-up and top-down processing as~\cite{wang2017residual} and~\cite{ronneberger2015u}, and is described as, 
\begin{align}
  \mathbf{A_{s}} = \phi(U(\mathbf{S})),
\end{align}
where $\phi$ is a nonlinear transform such as \textit{sigmoid} or \textit{softmax}, $U$ is a U-net like structure, composed of a series of downsampling and upsampling operations, and $\mathbf{S}$ is the input. As in~\cite{wang2017residual}, we used max-pooling for downsampling and bilinear interpolation for upsampling. In addition, skip connections between the corresponding bottom-up and top-down parts are added to help learn the attention weights. 

In contrast to~\cite{wang2017residual}, the attention weights in the model we propose are learned in the feature domain directly rather than in convolved domains. The motivation for this is twofold. First, the ASVspoof2017 dataset is much smaller than ImageNet and the approach of~\cite{wang2017residual} could underfit with the amount of training data available. Second, and more importantly, attention heatmaps at the input-feature level are much more readily interpretable than at subsequent convolution levels. 

\begin{figure}[tb]
 \centering
  \includegraphics[width=0.9\columnwidth]{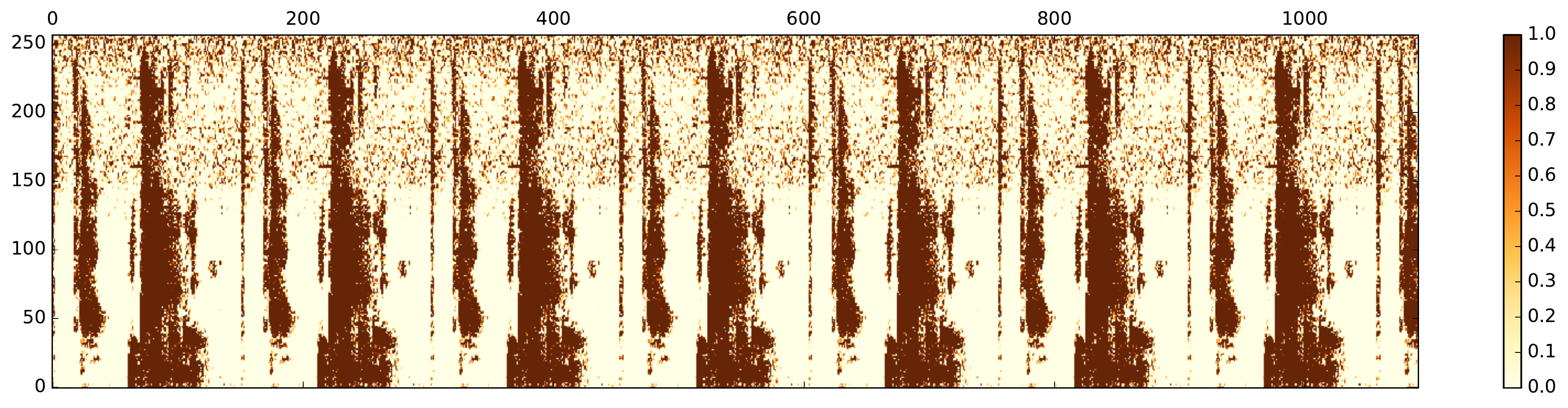}
  \includegraphics[width=0.9\columnwidth]{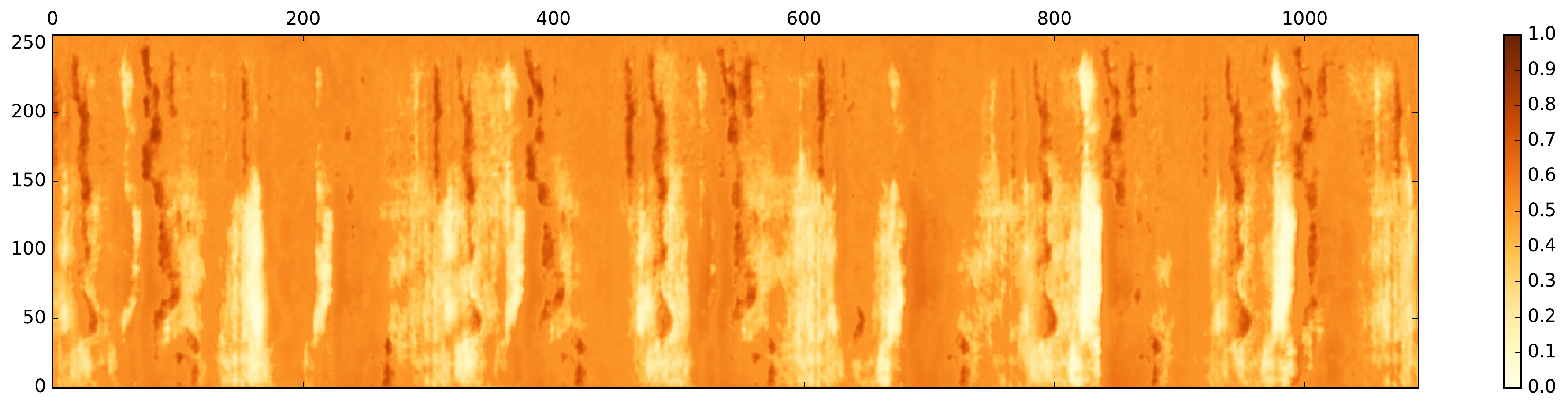}
  \includegraphics[width=0.9\columnwidth]{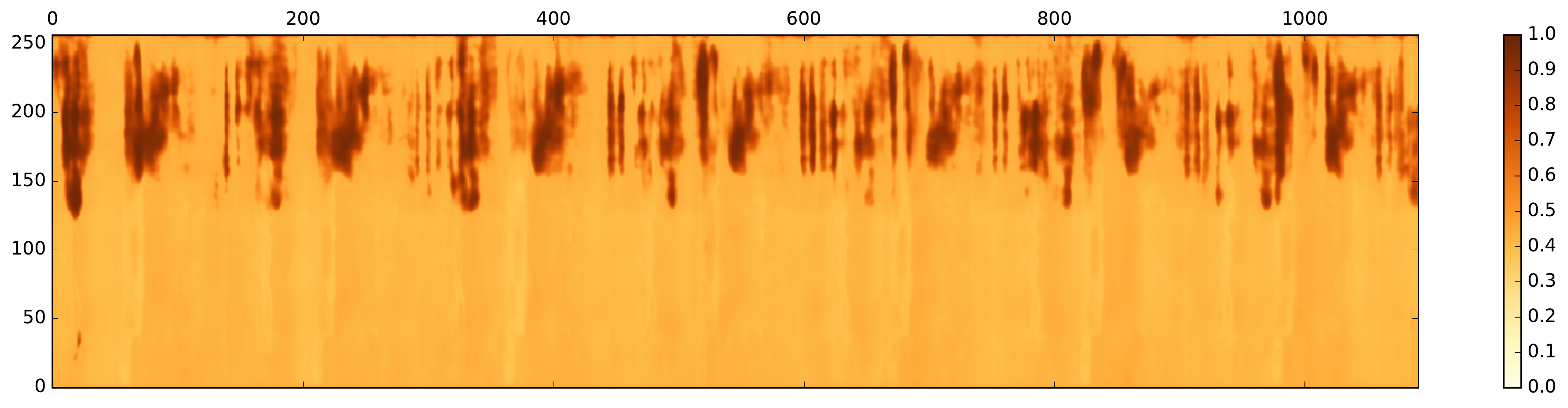}
\vspace{-3mm}
\caption{Visualizations of attention heatmaps: the original feature map (top) and the corresponding attention heatmaps learned without $\overline{\mathbf{S}}$ (middle) and  with $\overline{\mathbf{S}}$ (bottom).}
\label{fig:attention}
\vspace{-3mm}
\end{figure}

Figure \ref{fig:attention} shows attention heatmaps learned with and without $\overline{\mathbf{S}}$. As we can clearly see, the attention strongly focuses on high frequency components of speech segments when we trained the attention heatmaps with $\overline{\mathbf{S}}$, which is consistent with findings reported in literature. The intuitive explanation is that by giving DRN full information of $\mathbf{S}$, AF can focus on learning attention weights instead of learning a summary for $\mathbf{S}$. We can also see that with $\overline{\mathbf{S}}$, AF can selectively attend to and enhance not only high frequency segments but also any time and frequency segments. 

\subsection{Optimization}
AF and DRN are the two components of the Attentive Filtering Network, and the network is trained end-to-end. The Attentive Filtering Network is initialized with Xavier initialization~\cite{glorot2010understanding} and optimized with Adam with AMSGRAD~\cite{reddi2018convergence}. We also performed model selection based on equal error rates (EERs) measured on the development set after every training epoch, as we found this to yield better results. 


\section{Experiments}

\subsection{Baseline Systems}
\label{sec:baseline}

\noindent\textbf{CQCC-GMM:}
CQCCs~\cite{todisco2017constant} are derived using the constant Q transform, a perceptually motivated time-frequency analysis tool, and have been shown to be especially effective for spoofing countermeasures~\cite{todisco2017constant}. The baseline of the ASVspoof2017 Challenge uses CQCC features with a standard 2-class GMM classifier for genuine and spoofed speech~\cite{delgado2018asvspoof}. For each utterance the log-likelihood score is obtained from both models and the final system score is computed as the log-likelihood ratio.

\noindent\textbf{i-vector:}
The i-vectors~\cite{dehak2011front} pack a variable length speech recording into a fixed-dimension embedding. Following previous work~\cite{delgado2018asvspoof,adiban2017sut}, we experimented with a 64-mixture Universal Background Model (UBM) with 100-dimension i-vectors and a 128-mixture UBM with 200-dimension i-vectors. The i-vector extractors were trained on 30-dimensional CQCC features from the ASVspoof 2017 training set. We also length-normalized speaker-level i-vectors. The i-vectors were then averaged within each class, giving one i-vector representation for genuine speech and another i-vector representation for spoofed speech. We used two simple classifiers for the i-vectors: Gaussian linear generative model~\cite{martinez2011language} and cosine similarity~\cite{dehak2011front}. 

\noindent\textbf{LCNN:}
The best system submitted to the ASVspoof 2017 Challenge was based on the LCNN~\cite{lavrentyeva2017audio}, where a Max-Feature Map activation is used for a CNN~\cite{wu2018light}. We re-implemented the LCNN for audio replay attack detection according to the specification described in~\cite{lavrentyeva2017audio} and tested it on the ASVspoof 2017 Version 2.0 dataset.   

\subsection{Experimental Setup}
\label{sec:experiment}

All experiments in this work were conducted on Version 2.0 of the ASVspoof 2017 dataset \cite{delgado2018asvspoof}. The dataset has 1507 replay and 1507 \textit{bona fide} files in the training set, 950 replay and 760 \textit{bona fide} files in the development (\textit{dev}) set, and 12008 replay and 1298 \textit{bona fide} files in the evaluation (\textit{eval}) set. Details of the dataset can be found in~\cite{delgado2018asvspoof}. 
We used only the training set to train our system. The development set was used for model selection during validation and for tuning the logistic regression for system fusion. 
In this paper, we compare our proposed system with prior work on both Version 2.0 and Version 1.0 of the ASVspoof 2017 dataset.
Regarding implementation, the CQCC-GMM system was adopted from the official MATLAB code in~\cite{delgado2018asvspoof}. Log-spectrogram and i-vectors were extracted with Kaldi~\cite{povey2011kaldi}. Our Attentive Filtering Network was implemented in PyTorch, and the model implementation and training can be found in our github repository\footnote{\href{https://github.com/jefflai108/Attentive-Filtering-Network}{github.com/jefflai108/Attentive-Filtering-Network}}.

System fusion combines the strengths of different models to  improve overall performance. Fusion at the score level has been extensively used in ASVspoof2017 ~\cite{adiban2017sut,lavrentyeva2017audio,chen2017resnet}. In this work, we used the BOSARIS toolkit~\cite{brummer2011bosaris} for score fusion. The output of the network was first normalized using the \textit{dev} set. Then, logistic regression was conducted to derive the fusion weights and a bias. 

\subsection{Experimental Results of Single Systems}
\label{sec:results}

\begin{table}[tb]
\caption{EERs (\%) of the \textit{dev} and \textit{eval} sets of our single systems and published single and fusion systems. 
Parentheses after AF and DRN denote the nonlinear transforms used in AF and DRN, respectively.}
\label{tbl:EER}
\vspace{1mm}
\begin{tabular}{|p{37mm}|p{12mm}|p{12mm}|p{8mm}|}
\hline
Systems & \textit{dev} EER & \textit{eval} EER & Diff. \\ \hline
\textit{Version 2 dataset} & & & \\
AF(Sigmoid)-DRN(ReLU) & 6.55 & 8.99 & 2.44 \\ 
AF(SoftmaxT)-DRN(ReLU) & 6.62 & 9.28 & 2.66 \\ 
AF(SoftmaxF)-DRN(ReLU) & 6.52 & 9.34 & 2.82 \\ 
DRN(ELU) & 7.49 & 10.16 & 2.67 \\ 
AF(Tanh)-DRN(ReLU) & 6.87 & 10.17 & 3.30 \\ 
DRN(ReLU) & 6.69 & 10.30 & 3.61 \\ 
MDF(fusion)~\cite{suthokumar2018modulation} & - & \textbf{6.32} & - \\
qDFTspec~\cite{alam2018boosting} & - & 11.43 & - \\ 
CQCC-GMM(CMVN) \cite{delgado2018asvspoof} & 9.06 & 12.24 & 3.18 \\ 
i-vectors (Cosine Similarity) & 8.99 & 14.77 & 5.78 \\ 
i-vectors (Gaussian) \cite{martinez2011language} & 8.81 & 15.11 & 6.30 \\ 
LCNN (Our implementation) & \textbf{6.47} & 16.08 & 9.61 \\ 
Evolving RNN~\cite{valenti2018end} & 18.7 & 18.20 & -0.50 \\ 
CQCC-GMM & 12.08 & 29.35 & 17.27 \\ \hline
\textit{Version 1 dataset} & & & \\
DLFS(fusion)~\cite{saranyadecision} & 3.98 & \textbf{6.23} & 2.25 \\ 
MDF(fusion)~\cite{suthokumar2018modulation} & - & 6.54 & - \\
LCNN~\cite{lavrentyeva2017audio} & 4.53 & 7.34 & 2.81 \\ 
ConvRBM(fusion)~\cite{sailor2018auditory} & \textbf{0.82} & 8.89 & 8.07 \\
Multi-task~\cite{shim2018replay} & 4.21 & 9.56 & 5.35 \\
ResNet~\cite{chen2017resnet} & 10.95 & 16.26 & 5.31 \\\hline
\end{tabular}
\vspace{-5mm}
\end{table}

Table \ref{tbl:EER} compares various systems in terms of Equal Error Rates (EER) ($\%$). Overall, our proposed networks achieve competitive results on the Version 2.0 dataset.
%
The DRN system with ReLU activation function achieves $10.3$ EER on the \textit{eval} set, and the DRN with ELU activation function achieves $10.16$ EER.
By adding AF to the DRN, we can reduce EER further. Examining the effect of different non-linear activation functions, AF with \textit{Sigmoid} achieves $8.99$ EER, followed by  $9.28$ for \textit{SoftmaxT}, $9.34$ for \textit{SoftmaxF}, and $10.17$ for \textit{Tanh}. 
The Table also shows the absolute difference between EERs on the \textit{eval} set and \textit{dev} set. This indicates the extent to which the model may have been over-fitted to training samples. We see these differences for the proposed systems are small, indicating that the models generalize well even from the give small, unbalanced dataset.  

While~\cite{lavrentyeva2017audio} reported $7.37$ EER on the \textit{eval} set of  the Version 1.0 dataset, we could not replicate their results on the Version 2.0 dataset. With their LCNN, we obtained $16.08$ EER and the difference between the \textit{dev} and \textit{eval} sets is 9.61, which implies that LCNN could be over-fitting to the training data. Our i-vector baseline with cosine similarity backend achieves $14.77$ \textit{eval} EER, which is consistent with results reported in~\cite{delgado2018asvspoof}.


\subsection{Experimental Results using Fusion}

\begin{figure}[htb]
\vspace{-4mm}
  \centering
  \includegraphics[width=0.9\columnwidth]{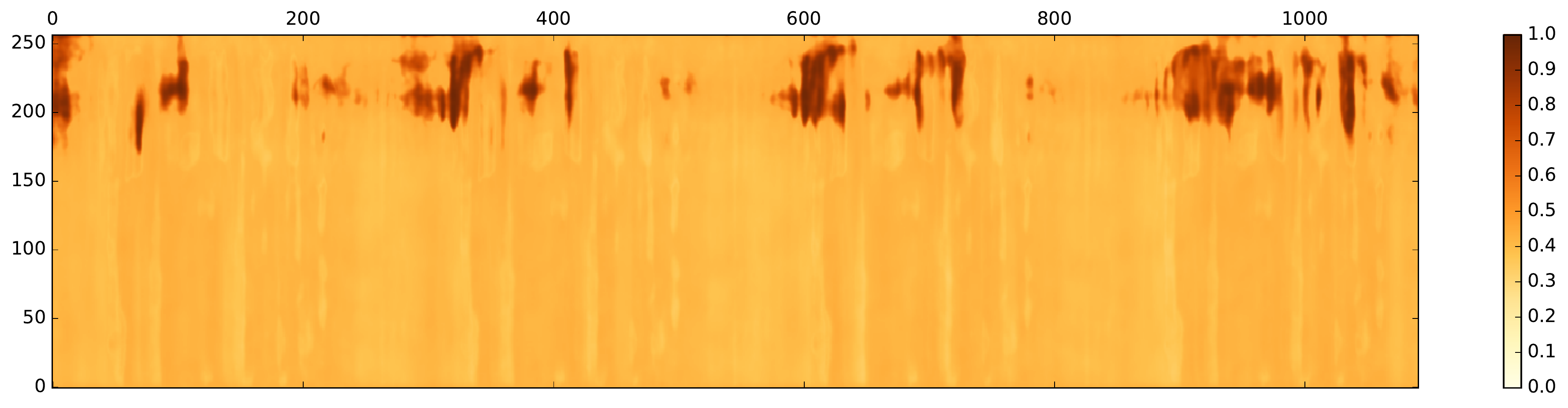}
  \includegraphics[width=0.9\columnwidth]{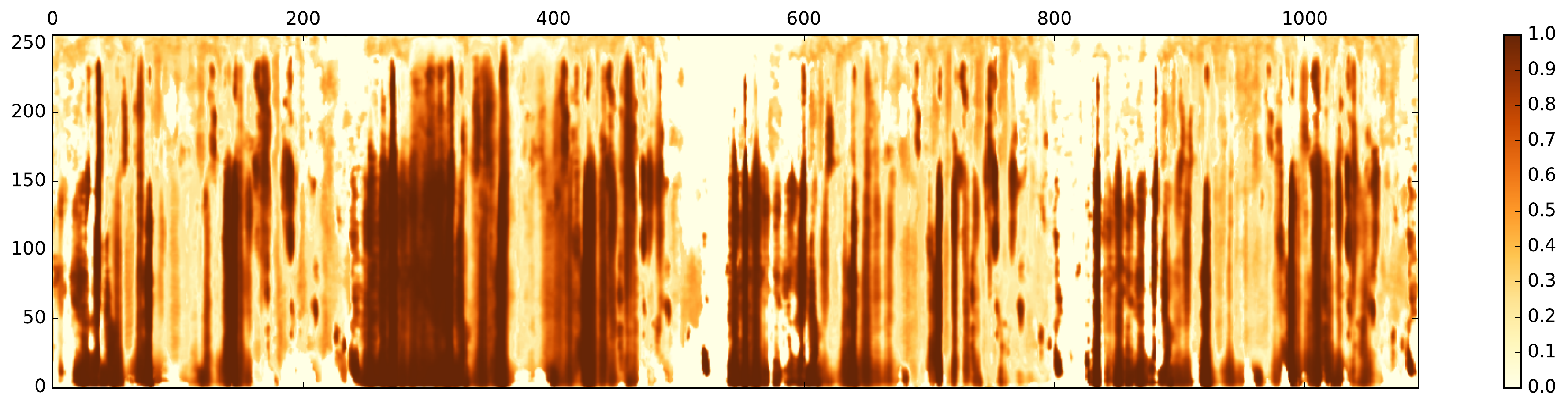}
  \includegraphics[width=0.9\columnwidth]{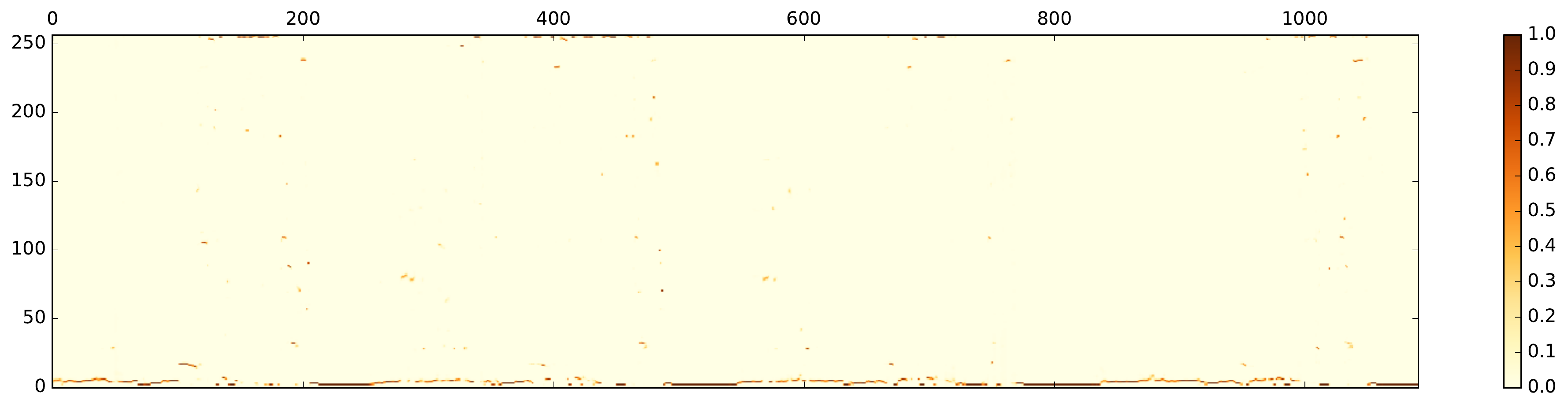}
  \includegraphics[width=0.9\columnwidth]{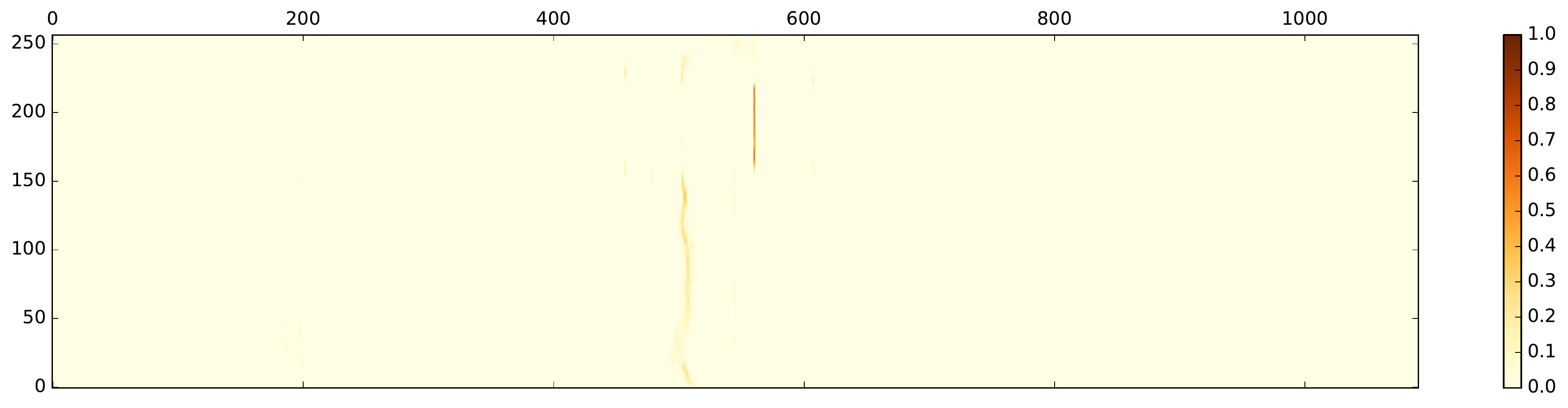}
\vspace{-4mm}
\caption{Visualizations of attention heatmaps corresponding to different nonlinearities (from top to bottom): \textit{Sigmoid}, \textit{Tanh}, \textit{SoftmaxF}, and \textit{SoftmaxT}.}
\label{fig:AF-activate}
\vspace{-8mm}
\end{figure}

The previous section showed that the performance of the proposed system varies depending on the nonlinear activation function used in the AF. We found this empirically because the learned attentions behave differently. Figure \ref{fig:AF-activate} shows attention heatmaps for different nonlinear transforms $\phi$ in Equation \ref{eq:attent}. We see that \textit{SoftmaxT} and \textit{SoftmaxF} enforce sparse activation, while \textit{Sigmoid} shows activations in multiple time and frequency bins. The difference in activations can be simply explained from how the nonlinearities scale each feature dimension. \textit{Sigmoid} scales each dimension independently while \textit{Softmax} scales each dimension dependently, implying that only a few dimensions are activated and most dimensions are suppressed (as shown in Figure \ref{fig:AF-activate} that most values are near 0). \textit{Softmax} works well for classification task but not in our context, where the scale of each dimension could be useful for detecting replay attacks.

Figure \ref{fig:AF-activate} also indicates that different nonlinear activation functions in AF adopt different aspects of the task, and since they may be complementary, we decided to fuse multiple AF systems using \textit{Sigmoid}, \textit{SoftmaxF} and \textit{SoftmaxT} activation functions. Table \ref{tbl:EER-fusion} gives the results of our fusion systems. As expected, the individual AF systems were complementary and fusing them reduced EER further. The best results for \textit{dev} and \textit{eval} were $6.09$ and $8.54$ respectively, obtained by fusing outputs of two AFNs with \textit{Sigmoid} and \textit{SoftmaxT} nonlinearity. 

\begin{table}[tb]
\caption{Fusion system results. Multiple AF systems using \textit{Sigmoid}, \textit{SoftmaxF} and \textit{SoftmaxT} were used to generate different attention maps. DRN(ReLU) was then used to calculate scores from each of the attention maps. We then fused the scores.}
\label{tbl:EER-fusion}
\vspace{1mm}
\centering
\begin{tabular}{|l|r|r|}
\hline
Fusion Systems & \textit{dev} EER & \textit{eval} EER \\ \hline
AF(Sigmoid)+AF(SoftmaxF) & 6.37 & 8.80 \\ \hline
AF(Sigmoid)+AF(SoftmaxT) & \textbf{6.09} & \textbf{8.54} \\ \hline
AF(SoftmaxT)+AF(SoftmaxF) & 6.39 & 8.98 \\ \hline
All & 6.29 & 8.67\\ \hline
\end{tabular}
\vspace{-5mm}
\end{table}

\section{Conclusions}
\label{sec:conclusion}
This paper presents our system for counteracting audio replay attacks. Our Attentive Filtering Network is composed of Attentive Filtering, which attends to and enhances input feature representation, and a Dilated Residual Network. Experiments conducted on the ASVspoof 2017 Version 2.0 dataset show the effectiveness of this model in replay attack detection, and furthermore, visualizing the attention heatmaps provides evidences for the network's feature enhancement behaviour. Our best single system achieved a competitive 8.99$\%$ evaluation EER, and our best fusion system provided 8.54$\%$ evaluation EER, providing a 30$\%$ relative improvement over the enhanced baseline system. 

\vspace{1mm}
\noindent
\textbf{Acknowledgments}
\label{sec:acknowledgment}
This work was done at the University of Edinburgh; the authors thank the CSTR group there for helpful discussions. CL was supported by the Johns Hopkins Vredenburg Scholarship. JY was supported by JSPS KAKENHI Grant Numbers 18KT0051 and by JST CREST Grant Number JPMJCR18A6.


\vfill\pagebreak

\bibliographystyle{IEEEbib}
\bibliography{main}

\end{document}